\documentclass[twocolumn,showpacs,preprintnumbers,amsmath,amssymb,aps]{revtex4}
\usepackage{graphicx}
\usepackage{bm}
\usepackage[mathscr]{eucal}
\usepackage[dvips]{color}

\begin{document}

\title{Interaction of thin tungsten and tantalum films with ultrashort laser pulses: calculations from first principles}

\author{N. A. Smirnov}

\email{deldavis@mail.ru}
\affiliation{Federal State Unitary Enterprise, Russian Federal Nuclear Center - Zababakhin All-Russian Research Institute of Technical Physics, 456770, Snezhinsk, Russia}
\date{\today}

\begin{abstract}
The interaction of ultrashort laser pulses with thin tungsten and tantalum films is investigated through the full-potential band-structure calculations. Our calculations show that at relatively low absorbed energies (the electron temperature $T_e$$\lesssim$7 kK), the lattice of tantalum undergoes noticeable hardening. The hardening leads to the change of the tantalum complete melting threshold under these conditions. Calculations suggest that for the isochorically heated Ta film, if such hardening really occurs, the complete melting threshold will be at least 25\% higher. It is also shown that the body-centered cubic structures of W and Ta crystals become dynamically unstable when the electronic subsystem is heated to sufficiently high temperatures ($T_e$$>$22 kK). This lead to their complete melting on the sub-picosecond time scale.
\end{abstract}

\maketitle

\section{Introduction}
As shown in a number of experimental studies, the melting of different materials after their interaction with ultrashort (femtosecond) laser pulses have their specific features \cite{A01,A02,A03,A04,A05}. Absorption of this radiation leads to a strongly non-equilibrium heating of the system where the temperatures of its electronic and ionic subsystems are very much different, $T_e$$\gg$$T_i$. This state may keep for tens of picoseconds and even longer \cite{A05}. Under these conditions, semiconductors, for example, undergo the so-called nonthermal melting caused not by their lattice heating due to heat transfer from hot electrons to cold ions but by a dramatic change in the shape of the potential energy surface and hence dynamic lattice destabilization \cite{A01,A02,A03}. In semimetallic bismuth, the situation seems to be similar \cite{A04}. The determining factor here is the estimate of the electron-phonon coupling factor $G$, which defines the rate of heat transfer from the electronic to ionic subsystem. For bismuth, the theoretical estimates of $G$ strongly differ \cite{A06,A07,A08}, leaving room for disputes on the presence of nonthermal melting in this metal after interaction with ultrashort laser pulses \cite{A09}.

On the other hand, the change of the shape of the potential energy surface may also lead, under certain conditions, to the hardening of irradiated crystal \cite{A10,A11,A12}, thus increasing the time of its melting and causing its strong overheating. Despite some claims that the lattice hardening has been experimentally observed \cite{A13}, there is still no evidence of its reliable detection in experiments \cite{A05,A12,A14}.

The experimental work reported in Ref.~\cite{A15} aimed to explore the possibility of the nonthermal melting of tungsten by measuring reflectivity of the metal surface after its irradiation. The experiments show that above a certain value of absorbed excitation fluence, ablation of the metal surface proceeds in a sub-picosecond time interval. The revealed effect may be indicative of the ultrafast nonthermal melting because in the normal thermal scenario of ablation, the characteristic times of this process must be much higher than those obtained in experiment \cite{A15}.

\emph{Ab initio} calculations \cite{A16} show that the heating of the electronic subsystem of tungsten to $T_e$ above 20 kK may lead to a structural transition from bcc to fcc phase. The transition is also caused by the abrupt change in the shape of the potential energy surface, leading to fcc stabilization at high values of $T_e$ \cite{A16}. In its turn, the bcc structure may lose dynamic stability under these conditions. It is however difficult to detect this transition in experiment because of the possibility of sub-picosecond nonthermal melting. Just this was shown in molecular dynamics (MD) calculations \cite{A17} where the interaction of femtosecond laser pulses with thin tungsten film was investigated. The nuclei of the new fcc phase were only able to form mainly on the surface of the film before the sample melted during about 0.8 ps. On whole, MD results \cite{A17} suggest that the detection probability for the nonthermal melting of tungsten is much higher than for the structural transition predicted in Ref.~\cite{A16}.

As mentioned above, an important factor of detecting nonthermal phenomena in metals is the electron-phonon coupling factor $G$. Its values for metals are usually high \cite{A18}, meaning that the nonthermal character of processes that occur after irradiation can hardly be recognized. There are different approaches to the theoretical determination of $G$ (see, for example, \cite{A12,A18,A19}). In our research we will follow methodology described in Ref.~\cite{A12}, but also discuss results obtained with other approaches.

This paper studies the interaction of femtosecond laser pulses with thin (a few tens of nanometers thick) tungsten and tantalum films. The physical quantities required for calculations with a two-temperature model \cite{A20} were obtained from first principles. The issues discussed include the processes involved in the nonthermal melting of the metals and the possibility of detecting tantalum lattice hardening at moderate absorbed energies. Our results are compared with available experimental data and other calculations.

\section{Calculation method}
In this work, the temperature evolution of electronic and ionic subsystems with time after irradiation by ultrashort laser pulses is determined using a well-known two-temperature model \cite{A20}. Since the thin ($\sim$10 nm) films of W and Ta are considered, the two-temperature model equations can be written as
\begin{equation}
C_e(T_e)\frac{\partial T_e}{\partial t}=-(T_e-T_i)G(T_e)+S(t),\label{eq01}
\end{equation}
\begin{equation}
C_i(T_i)\frac{\partial T_i}{\partial t}=(T_e-T_i)G(T_e),\label{eq02}
\end{equation}
where $S$($t$) is the time dependent radiation source function \cite{A17}, $C_e$($T_e$) and $C_i$($T_i$) are electron and lattice heat capacities, and $G$($T_e$) is the electron-phonon coupling factor. Here we neglect lattice ($\kappa_i$) and electron ($\kappa_e$) thermal conductivities because, on the one hand, $\kappa_e$$\gg$$\kappa_i$ in our case, and on the other hand, in thin foils, ballistic electrons bring the electronic subsystem to thermodynamic equilibrium over a time about a pulse duration $\tau_p$ \cite{A21,A22}. So, no significant gradients in temperature occur in the target. The method to calculate $C_e$, $C_i$, and $G$ as functions of electron and ion temperatures from first principles is described in rather detail in Ref.~\cite{A12}. Here we only provide the key formula for the electron-phonon coupling factor. It reads as
\begin{multline}
G(T_e)=\frac{2\pi\hbar}{(T_l-T_e)}\int\limits_0^{\infty}\Omega{d\Omega}\int\limits_{-\infty}^{\infty}N(\varepsilon)\alpha^2F(\varepsilon,\Omega)\\
\times S(\varepsilon,\varepsilon+\hbar\Omega)d\varepsilon. \label{eq03}
\end{multline}
where $N$($\varepsilon$) is the electronic density of states (DOS), $\alpha_2F$($\varepsilon$,$\Omega$) is the electron-phonon spectral function, $\varepsilon$ and $\hbar\Omega$ are, respectively, electron and phonon energies, $S$($\varepsilon$,$\varepsilon+\hbar\Omega$)$=$[$f_e$($\varepsilon$)-$f_e$($\varepsilon+\hbar\Omega$)][$n$($\hbar\Omega$,$T_i$)-$n$($\hbar\Omega$,$T_e$)] with $f_e$ standing for the Fermi distribution function and $n$ for the Bose-Einstein distribution function.

Another formula which is often used to determine $G$($T_e$) has some simplifications as compared to (\ref{eq03}) and reads as \cite{A18}
\begin{equation}
G(T_{e})=\frac{\pi\hbar{k_{B}}\lambda\langle\omega^2\rangle}{N(E_{F})}\int\limits_{-\infty}^{\infty}N^2(\varepsilon)\left(-\frac{\partial{f_e}}{\partial\varepsilon}\right)d\varepsilon. \label{eq04}
\end{equation}
Here $\lambda$ is the electron-phonon mass enhancement parameter, $\langle\Omega\rangle^2$ is the second moment of the phonon spectrum \cite{A23}, and $E_F$ is the Fermi energy. Formula~\ref{eq04} is derived under the assumption that in the interaction with phonon, the scattering probability matrix elements is independent of the initial $\{\textbf{k},i\}$ and final {$\{\textbf{k}',j\}$} electronic states. The authors of Ref.~\cite{A18} determined the values of $\lambda$ and $\langle\Omega\rangle^2$ from experimental evaluation, not from first-principles calculations.

One more way to calculate $G$($T_e$) is based on the calculation of the electron-ion collision integral $I_{nm}^{e-i}$ with the use of an approximate tight-binding model to calculate the band structure, combined with MD simulation \cite{A19}. The expression for $I_{nm}^{e-i}$ is written as
\begin{widetext}
\begin{equation}
I_{nm}^{e-i}=\frac{2\pi}{\hbar}|M_{e-i}(\varepsilon_n,\varepsilon_m)|^2
\begin{cases}
f_e(\varepsilon_n)[2-f_e(\varepsilon_m)]-f_e(\varepsilon_m)[2-f_e(\varepsilon_n)]e^{-\Delta\varepsilon/T_i};&\text{for $n$$>$$m$}\\
f_e(\varepsilon_m)[2-f_e(\varepsilon_n)]e^{-\Delta\varepsilon/T_i}-f_e(\varepsilon_n)[2-f_e(\varepsilon_m)];&\text{otherwise}
\end{cases}, \label{eq05}
\end{equation}
\end{widetext}
where $\Delta\varepsilon$$=$$\varepsilon_n-\varepsilon_m$ is the energy difference between two states, and $M_{e-i}$ is the electron-ion scattering matrix element. The electron-phonon coupling factor can be written as
\begin{equation}
G(T_e)=\frac{1}{V(T_e-T_i)}\sum_{n,m}\varepsilon_m I_{nm}^{e-i}, \label{eq06}
\end{equation}
here $V$ is the specific volume. It should be noted here that our method for determining $G$($T_e$) (by formula (\ref{eq03})) does not use any experimentally determined parameters or approximations which simplify the scattering probability matrix element, as it is done in Ref.~\cite{A18}, or serious simplifications related to particle interactions in the system, as it is done in the tight-binding model \cite{A19}.

In this work, first-principles calculations were done with the all-electron full-potential linear muffin-tin orbital method (FP-LMTO) \cite{A24}. We consider here processes at a constant specific volume, i.e. the isochoric heating of targets. Within the scope of density functional theory the FP-LMTO method calculates the electron structure, internal and free energies, phonon spectrum and other material properties \cite{A12,A24,A25,A26}. Phonon spectrum and electron-phonon spectral function calculations for the metals of interest were done with linear response theory implemented in the FP-LMTO code \cite{A24,A25}. Integration over the Brillouin zone was done with an improved tetrahedron method \cite{A27}. Meshes in $\textbf{k}$-space corresponded to equidistant spacing 30$\times$30$\times$30. For integration over the $\textbf{q}$-points of the phonon spectrum, a 10$\times$10$\times$10 mesh appeared quite sufficient (see \cite{A26} for more details on meshes). The cutoff energy for representing the basis functions as a set of plane waves in the interstitial region was taken to be 900 eV. The basis set included MT-orbitals with moments to $l^b_{max}$$=$5. Charge density and potential expansions in terms of spherical harmonics were done to $l^w_{max}$$=$7. The internal FP-LMTO parameters such as the linearization energy, tail energies, and the radius of the MT-sphere were chosen using an approach similar to that one used in Ref.~\cite{A28}.
\begin{figure}
\centering{
\includegraphics[width=8.0cm]{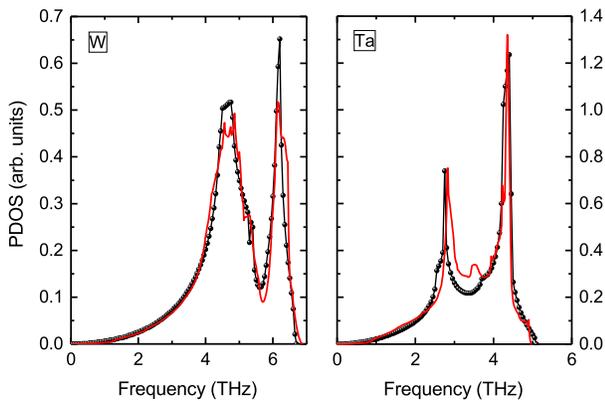}}
\caption{\label{fig1} Tungsten and tantalum phonon spectra at the equilibrium experimental specific volume from calculations done in this work for zero temperature (red lines) and from experiment at room temperature \cite{A30} (circles connected by a line).}
\end{figure}

The valence electrons in our calculations were 5$s$, 5$p$, 4$f$, 5$d$, and 6$s$. For better comparison with calculations by other authors, the exchange-correlation potential was chosen to be similar to that one used in Ref.~\cite{A17}, i.e., PBE \cite{A29}. This functional reproduces well the different properties of tungsten and tantalum. For example, the equilibrium volume $V_0$ from calculation differs by no more than 2\% from experiment for both the metals. Figure~\ref{fig1} shows the phonon densities of states (PDOS) from calculation in comparison with experimental data \cite{A30}. They are seen to be in quite a good agreement.

The entropy of the electronic subsystem was determined as
\begin{equation}
S_e(T_e)=-k_B\int^{\infty}_{-\infty}d\varepsilon N(\varepsilon)[f_e ln(f_e)+(1-f_e)ln(1-f_e)]. \label{eq07}
\end{equation}
With the known entropy $S_e$($T_e$) and internal energy $E_e$($T_e$) of electrons, it is easy to obtain the free energy $F_e$$=$$E_e-T_{e}S_{e}$ of the electron gas.

The phonon spectrum of tungsten and tantalum was determined within quasiharmonic approximation \cite{A12}. The melting temperature $T_m$ of crystal W and Ta versus electron temperature was estimated in the same manner as it was done in Ref.~\cite{A31} with the well performing Lindemann criterion.

\section{RESULTS}
\begin{figure}
\centering{
\includegraphics[width=8.0cm]{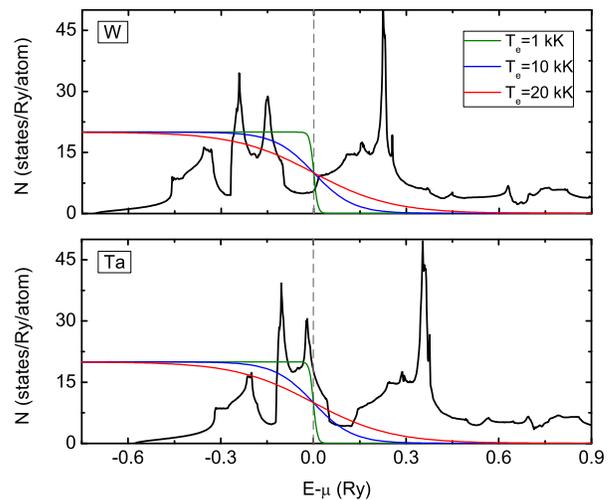}}
\caption{\label{fig2} Electronic DOS for W (top) and Ta (bottom) at equilibrium specific volume and zero temperature (black lines). The green, blue and red lines are the Fermi distribution functions at different electron temperatures.}
\end{figure}

Let's first compare the electronic structures of tungsten and tantalum. Figure~\ref{fig2} shows their electronic densities of states versus energy at $V$$=$$V_0$ and $T$$=$0 calculated in this work. It is seen that the chemical potential µ which coincides with the Fermi energy at zero temperature is near the minimum of the DOS for tungsten, while for tantalum, the density of states at $\varepsilon$$=$$\mu$ is much higher compared to W. For Ta, the Fermi level is near the peak of the DOS. Compared to tantalum, the electronic structure of tungsten is very much depleted in states in the vicinity of $\mu$. Calculations show that as $T_e$ grows to $\sim$15 kK, the values of $N$($\mu$) increase for tungsten and decrease for tantalum. This causes certain differences in the behavior of these metals at elevating electron temperatures.

\begin{figure}
\centering{
\includegraphics[width=8.0cm]{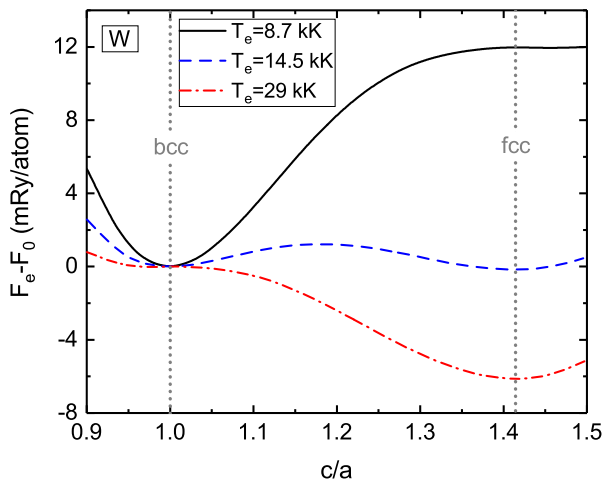}}
\caption{\label{fig3} Free electron energy versus lattice parameter $c/a$ at different $T_e$ for tungsten ($V$$=$$V_0$). The vertical lines show the values of $c/a$ which correspond to its bcc and fcc structures.}
\end{figure}

\begin{figure}
\centering{
\includegraphics[width=8.0cm]{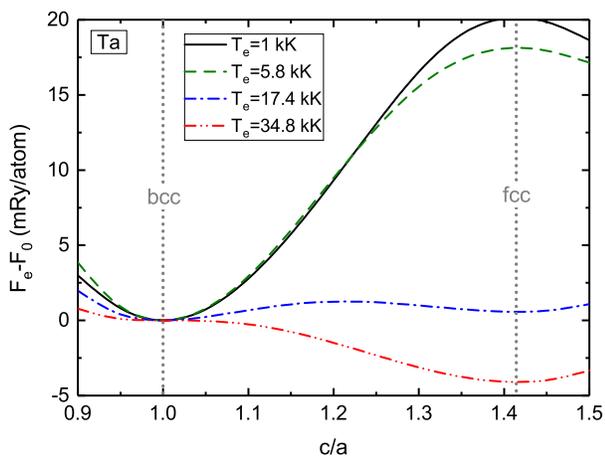}}
\caption{\label{fig4} Free electron energy versus lattice parameter $c/a$ at different $T_e$ for tantalum ($V$$=$$V_0$). The vertical lines show the values of $c/a$ which correspond to its bcc and fcc structures.}
\end{figure}

Now consider how the free energy of electrons depends on the lattice parameter $c/a$ (i.e., the Bain path) at different temperatures $T_e$. Figures~\ref{fig3} and \ref{fig4} show results obtained for W and Ta, respectively. In both metals, the fcc structure is seen to be dynamically unstable at low electron temperatures. With the increasing temperature it stabilizes and at $T_e$$>$15 kK it becomes thermodynamically more preferable than bcc. It is seen that tantalum behaves very much like tungsten but requires somewhat higher temperatures for stabilization of the fcc structure. On the other hand, with the increasing $T_e$ the bcc structure becomes dynamically unstable both in tungsten and in tantalum. These changes must lead to a bcc$\rightarrow$fcc transition when the electronic subsystem is heated. As however mentioned in paper \cite{A17}, in such conditions their melting is more probable. On whole, our calculations for tungsten agree well with results presented in Ref.~\cite{A16}.

\begin{figure}
\centering{
\includegraphics[width=8.0cm]{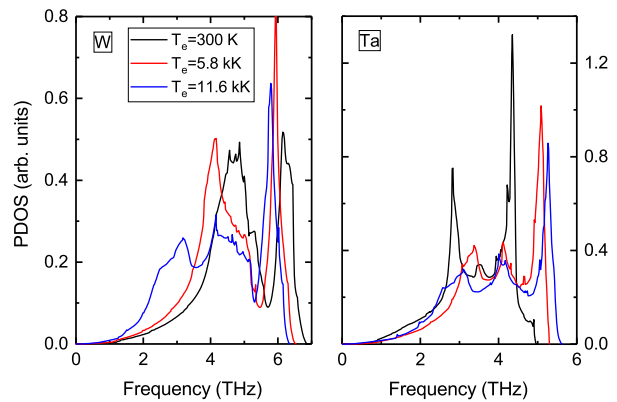}}
\caption{\label{fig5} Phonon densities of states in tungsten (left) and tantalum (right) at different electron temperatures ($V$$=$$V_0$).}
\end{figure}

One more feature of tantalum should be noted here. It is seen from Fig.~\ref{fig4} that there exists a limited interval of temperatures at relatively low values of $T_e$ (see $T_e$$=$5.8 kK), where the bcc lattice hardens. The free energy curve runs steeper near the minimum corresponding to the bcc phase. This feature is absent in tungsten. Figure~\ref{fig5} shows the densities of phonon states for W and Ta we calculated in this work for different electron temperatures. It is seen that with the increasing $T_e$ tungsten gradually softens and its phonon frequencies reduce. The phonon frequencies of tantalum first increase with the growing $T_e$ and cause bcc lattice hardening. Then the tendency changes – the high-frequency part of the spectrum goes on to harden, while the low-frequency part begins to soften reducing its frequencies (see Fig.~\ref{fig5}, $T_e$$=$11.6 kK). At $T_e$ above 20 kK the bcc structure in both metals loses its dynamic stability. It happens at about 22 kK in tungsten and 29 kK in tantalum. The hardening of the Ta lattice at relatively low electron temperatures leads to a sudden effect we will consider later.

\begin{figure}
\centering{
\includegraphics[width=8.0cm]{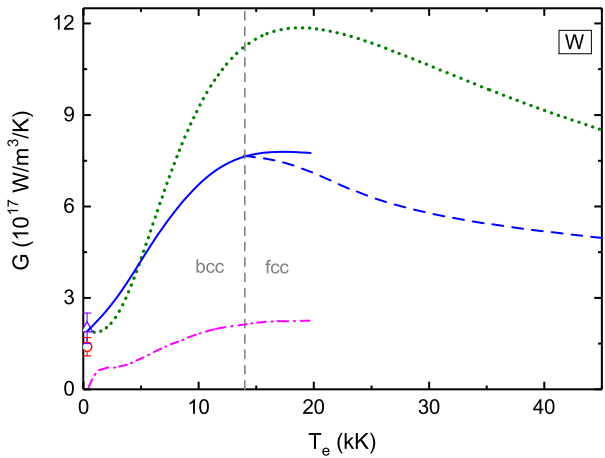}}
\caption{\label{fig6} Electron-phonon coupling factor versus $T_e$ for tungsten from our calculation (solid, dashed lines for bcc and fcc, respectively), from calculations reported in papers \cite{A18} (dotted line) and \cite{A19} (dashed-dotted line), and from experiments \cite{A32} and \cite{A33} (the circle and the triangle, respectively). The vertical line shows the approximate value of $T_e$ above which the fcc phase becomes more energetically favorable than bcc.}
\end{figure}

\begin{figure}
\centering{
\includegraphics[width=8.0cm]{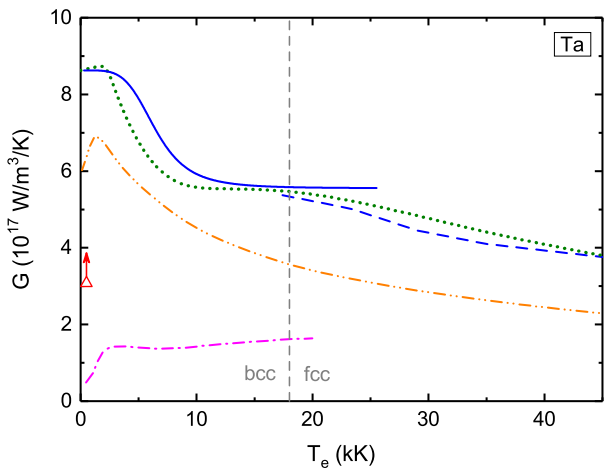}}
\caption{\label{fig7} Electron-phonon coupling factor versus $T_e$ for tantalum from our calculations using formula~(\ref{eq03}) (solid, dashed lines for bcc and fcc, respectively) and by a formula~(\ref{eq04}) (dotted line). Other calculations: dashed-dotted line - Ref.~\cite{A19}, dashed-dotted-dotted line - Ref.~\cite{A34} by a formula from Ref.~\cite{A18} (see the text). The triangle shows the lower boundary of $G$ from experiment \cite{A34}. The vertical line shows the approximate value of $T_e$ above which the fcc phase is more energetically preferable than bcc.}
\end{figure}

Figures~\ref{fig6} and \ref{fig7} show the electron-phonon coupling factor $G$ as a function of electron temperature at $V$$=$$V_0$, calculated in this work for tungsten and tantalum, respectively. The dependences $G$($T_e$) are provided for bcc and fcc structures in their stability regions. The values of G for the structures are seen to be close to each other and it is quite possible to approximate our results by a continuous line. The figures also show data from low-temperature experiments \cite{A32,A33,A34}. For tungsten, our results are seen to agree quite well with experiment. For tantalum, experimental data from Ref.~\cite{A34} provides only the lower boundary of $G$, which does not contradict our calculations. Figures~\ref{fig6} and \ref{fig7} also show results from some other calculations. It is seen that compared to our results, calculations by Lin et al.~\cite{A18} for W give overestimated values of $G$ for the increasing temperature (Fig.~\ref{fig6}). Such a behavior has earlier been observed in other metals \cite{A12} and can be related to the more correct account for the energy dependence of $\alpha_2F$($\varepsilon$,$\Omega$) in formula~(\ref{eq03}). In turn, the values of $G$($T_e$) from Ref.~\cite{A19} are much lower than our results and the experimental data available. Note that the presence of adjustable parameters in the calculation method may reduce the accuracy of results if they are adjusted to conditions (for example, at $T$$=$0) different from what we are having here.

For tantalum (fig.~\ref{fig7}), our calculations by expression~(\ref{eq04}) (the dotted line) had one distinction from those reported in paper~\cite{A18}: the values of $\lambda$ and $\langle\Omega\rangle^2$ were determined from first-principles calculations rather than from experimental evaluation. It is seen that in this case, approaches \cite{A18} and \cite{A12} give close values for $G$($T_e$), the differences are minimal. In Ref.~\cite{A34}, the electron-phonon coupling factor was also calculated with formula~(\ref{eq04}) but with the electronic DOS determined from MD calculations. But here deviations from our results come, first of all, from the underestimated parameter $\lambda$. The authors of \cite{A34} used the empirical value from Ref.~\cite{A23}, $\lambda$$=$0.65. Our calculations from first principles gave $\lambda$$=$0.88 in the case of tantalum. For tungsten, the difference between the empirical \cite{A23} and calculated values of $\lambda$ is not so large; they agree within $\sim$3\%.
\begin{figure}
\centering{
\includegraphics[width=8.0cm]{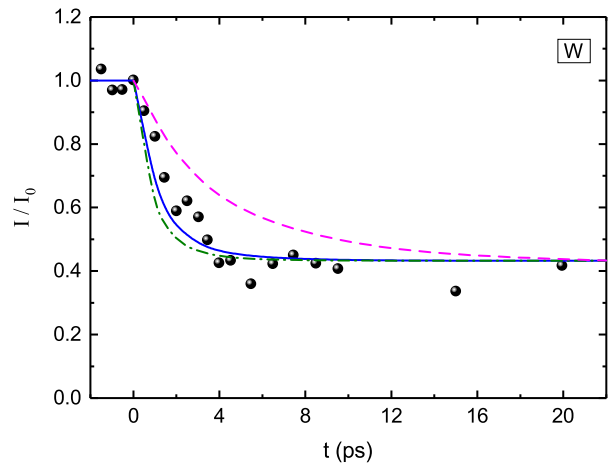}}
\caption{\label{fig8} Intensity of diffraction peak (211) versus time for tungsten for absorbed energy density 0.8 MJ/kg from our calculation (the solid line), calculations with a constant $G$ \cite{A33} (the dashed line), calculations with $G$($T_e$) from Ref.~\cite{A18} (the dashed-dotted line), and measurements \cite{A33} (circles).}
\end{figure}

Let's consider the accuracy of our calculations in comparison with other experimental results. The authors of paper~\cite{A33} measured how evolved the intensity of the Laue diffraction peak (211) after a 30-nm-thick tungsten film deposited on a silicon nitride substrate was irradiated by 400-nm laser pulses with $\tau_p$$=$130 fs. The absorbed energy density $E_{abs}$ was about 0.8 MJ/kg. Figure~\ref{fig8} compares experimental data with calculations performed in three variants (see \cite{A12} for calculation details). In addition to our computation with use of formula~(\ref{eq04}), it shows calculations with $G$($T_e$) taken from Ref.~\cite{A18} and with constant $G$$=$$2\cdot10^{17}$ $\text{W}/\text{m}^3/\text{K}$ and $\Theta_D$$=$312 K \cite{A33}. The results obtained with expression~(\ref{eq03}) are seen to agree quite well with experiment. The use of $G$($T_e$) from Ref.~\cite{A18} slightly worsens the agreement and the calculation with the constant $G$ markedly underestimates the change of the diffraction peak intensity at times below 10 ps.

\begin{figure}
\centering{
\includegraphics[width=8.0cm]{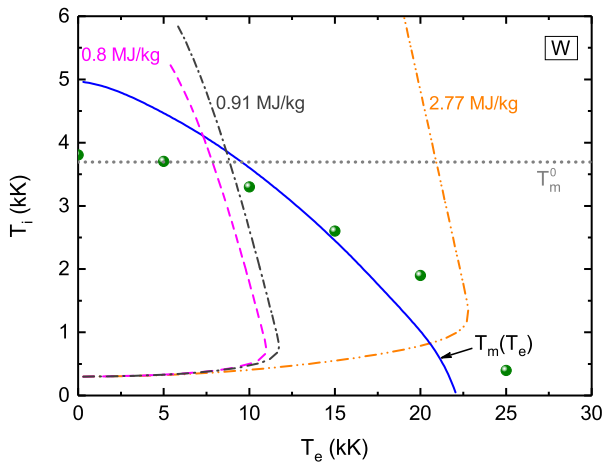}}
\caption{\label{fig9} Calculated evolution of electron and ion temperatures (isochoric heating) after irradiation of the 30-nm-thick tungsten film by a 130-fs pulse for different absorbed energy densities (dashed, dashed-dotted, and dashed-dotted-dotted lines). The solid line shows the melting temperature $T_m$ as a function of $T_e$ from our calculation, the circles show $T_m$($T_e$) from Ref.~\cite{A17} (non-isochoric conditions), and the dotted line shows the normal melting temperature of W.}
\end{figure}

Figure~\ref{fig9} presents ion temperature versus electron temperature for tungsten, calculated by solving equations (\ref{eq01})-(\ref{eq02}). We reproduced experimental conditions from Ref.~\cite{A33} but did calculations for several values of $E_{abs}$. The possibility of the bcc$\rightarrow$fcc transition was not considered because ultrafast melting was here more probable \cite{A17}. Figure~\ref{fig9} also shows the melting temperature of W versus $T_e$, obtained in this work and by Murphy et al. \cite{A17} from MD calculations. Remind that our $T_m$($T_e$) was calculated with the Lindemann criterion. As seen from Fig.~\ref{fig9}, the melting temperature of tungsten decreases with the increasing $T_e$ due to lattice softening (Fig.~\ref{fig5}). The resulted dependence $T_m$($T_e$) agrees rather well with data from Ref.~\cite{A17} despite the essentially different approaches to its determination. Some discrepancy comes from the fact that our calculation corresponded to the isochore $V$$=$$V_0$, while in MD simulation \cite{A17}, the sample could expand along the axis normal to the target surface.

In paper~\cite{A33}, a threshold value $E^m_{abs}$ required for the complete melting of tungsten was determined. For the conditions of that experiment, it was found to be 0.9 MJ/kg. Our calculations give a very close value of 0.91 MJ/kg (details of calculation can be found in paper \cite{A12}). Complete melting occurs after the temperature $T_m$ is reached and the lattice gets sufficient heat to overcome the latent heat of fusion, $\Delta H_m$ \cite{A35}. The absorbed energy density of 0.8 MJ/kg is not enough to completely melt the target \cite{A33}. It is seen from Fig.~\ref{fig9} that at high $E_{abs}$ ($>$2.5 MJ/kg) the lattice temperature $T_i$ reaches $T_m$ even earlier than $T_i$($T_e$) reaches its maximum. At high $T_e$, the melting temperature of tungsten becomes much lower than the normal melting temperature determined at ambient pressure, $T^0_m$$\approx$3.7 kK. MD calculations and analytic equations of state \cite{A36,A37}, including that one for tungsten, suggest that the heat of fusion changes under the action of external conditions and it will reduce as $T_m$ decreases. This will also influence the time of melting. Usually, $T_e$ reaches a maximum after irradiation by ultrashort pulses at a time of about a few $\tau_p$. Therefore at sufficiently high $E_{abs}$ ($>$2.5 MJ/kg) tungsten will melt during sub-picosecond times which is also proved by calculations \cite{A17}.
\begin{figure}
\centering{
\includegraphics[width=8.0cm]{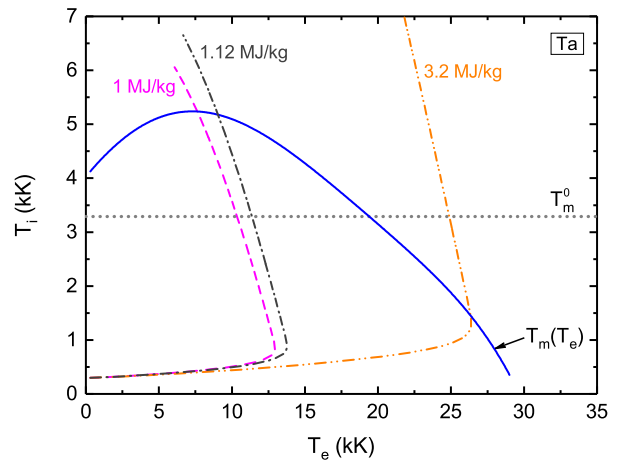}}
\caption{\label{fig10} Calculated evolution of electron and ion temperatures (isochoric heating) after irradiation of a 30-nm-thick tantalum film by a 130-fs-pulse for different absorbed energy densities (dashed, dashed-dotted, and dashed-dotted-dotted lines). The solid line shows $T_m$ versus $T_e$ from our calculation and the dotted line shows the normal melting temperature of Ta.}
\end{figure}

Now consider tantalum. Figure 10 demonstrates the $T_i$($T_e$) dependence for Ta similarly to tungsten. Irradiation conditions and target thickness are the same as for W. It is seen that the melting curve $T_m$($T_e$) reaches a maximum approximately at $T_e$$=$7.3 kK due to the hardening of the Ta crystal lattice at these temperatures, as mentioned earlier (see Fig.~\ref{fig5}). Unlike gold, whose melting temperature begins to increase only at $T_e$$>$15 kK (remaining almost constant at lower $T_e$) \cite{A12}, for tantalum this growth of $T_m$ starts right after the electron temperature increases. At $T_e$ higher than 7.3 kK, its lattice begins to gradually soften. Like tungsten, tantalum at sufficiently high values of $E_{abs}$ ($>$3 MJ/kg) must melt on the sub-picosecond time scale due to the loss of dynamic stability by its lattice (Fig.~\ref{fig10}). We do not consider the bcc$\rightarrow$fcc transition here also. The high electron-phonon coupling factor of tantalum signals a higher probability of its ultrafast melting. However, the existence of a maximum of $T_m$($T_e$) at relatively low electron temperatures gives an interesting effect. If such hardening really occurs, it should lead to an increase in the melting threshold $E^m_{abs}$ for Ta metal. As shown in calculations, $E^m_{abs}$ will be at least 25\% higher. For tantalum normal melting temperature, $T^0_m$$=$3.29 kK, the threshold value $\widetilde{E}^m_{abs}$ equals 0.74 MJ/kg. If the crystal lattice hardens, then, under isochoric heating, an absorbed energy density of $\sim$1.12 MJ/kg is required for complete melting. For non-isochoric conditions, the threshold may be lower, about 0.93 MJ/kg. However, the value is still rather far from normal $\widetilde{E}^m_{abs}$=0.74 MJ/kg and can be determined quite reliably in experiment (see, for example, \cite{A05}). In addition, the growth of $T_m$ make the latent heat of fusion higher which will also delay the complete melting.

A similar maximum of $T_m$($T_e$) at relatively low heating ($T_e$$\sim$5 kK) is also present in platinum \cite{A12}. As shown by calculations from first principles, its electronic structure is also characterized by a high electronic density of states $N$($\mu$) on the Fermi level \cite{A18}, which strongly reduces with the increasing $T_e$. Our calculations show that the effect of lattice hardening is a bit lower here and the melting threshold increases by about 18\%. But since $\widetilde{E}^m_{abs}$ for platinum at the normal melting temperature $T^0_m$ is quite small ($\sim$0.39 MJ/kg), the detection of its increase in experiment may be limited by experimental accuracy.

\section{Conclusions}
The paper studied the interaction of femtosecond laser pulses with thin tungsten and tantalum films through calculations from first principles. Calculated results shows the body-centered cubic structure of both the metals to lose its dynamic stability at rather high electron temperatures. This effect must lead to their melting on the sub-picosecond time scale when the electronic subsystem is heated above 22 kK. It is also demonstrated that the metals have rather high values of the electron-phonon coupling factor ($\sim$ several units per $10^{17}$ $\text{W}/\text{m}^3/\text{K}$) at electron temperatures from room temperature to $\sim$45 kK. In addition, unlike tungsten, the crystal lattice of tantalum hardens at relatively low values of $T_e$ ($\lesssim$7 kK). The hardening changes the value of the complete melting threshold. Our calculations show that the melting threshold will be at least 25\% higher if hardening really occurs. We suppose that this effect for tantalum can be detected quite reliably by modern experimental techniques used to study the interaction of matter with ultrashort laser pulses.

\end{document}